\documentclass[DIV12]{scrartcl}
\usepackage{etex}
\usepackage{flushend}

\usepackage[utf8]{inputenc}
\usepackage{xspace}
\usepackage{amsfonts}
\usepackage{mathptmx}
\usepackage{amssymb}
\usepackage{mathtools}
\usepackage{xcolor}
\usepackage{paralist}
\usepackage{booktabs}
\usepackage{tabularx}
\usepackage{wrapfig}
\usepackage{needspace}
\usepackage[pdfborder={0 0 0}]{hyperref}

\usepackage{tikz}
\usepackage{tikz-qtree}

\usepackage{stmaryrd}

\newtheorem{theorem}{Theorem}

\title{Structurally Tractable Uncertain Data}
\author{Antoine Amarilli\\Institut Mines--Télécom\\Télécom ParisTech; CNRS LTCI}
\date{}

\makeatletter
\newcommand*{\defeq}{\mathrel{\rlap{%
  \raisebox{0.3ex}{$\m@th\cdot$}}%
  \raisebox{-0.3ex}{$\m@th\cdot$}}%
  =}
\makeatother

\renewcommand{\phi}{\varphi}
\renewcommand{\epsilon}{\varepsilon}

\newcommand\restr[2]{{
  \kern-\nulldelimiterspace 
  #1 
  _{|#2} 
  }}

\newcommand{\prxml}{\mathsf{PrXML}}

\newcommand{\mux}{\prxmlclass{mux}}
\newcommand{\ind}{\prxmlclass{ind}}

\newcommand{\prxmlclass}[1]{\mathsf{#1}}

\newcommand{\cie}{\prxmlclass{cie}}

\newcommand{\acz}{\mathrm{AC}^{\smash{0}}}

\begin{document}
\maketitle

\begin{abstract}
  Many data management applications must deal with data which is uncertain,
incomplete, or noisy. However, on existing uncertain data representations, we
cannot tractably perform the important query evaluation tasks of determining
query possibility, certainty, or probability: these problems are hard on arbitrary
uncertain input instances. We thus ask whether we could restrict the structure
of uncertain data so as to guarantee the tractability of exact query evaluation. We
present our tractability results for tree and tree-like uncertain data, and a
vision for probabilistic rule reasoning. We also study uncertainty about order,
proposing a suitable representation, and study uncertain data conditioned by
additional observations.

\end{abstract}

\section{Introduction}
\label{sec:introduction}
Traditional database management theory assumes that
data is correct and complete.
However, more and more applications deal with incomplete,
uncertain, and noisy data. For instance, data is extracted or inferred automatically
from random 
Web pages by automated and error-prone extraction
programs~\cite{carlson2010toward}; 
integrated from diverse sources 
through approximate mappings~\cite{dong2007data};
contributed
to collaboratively editable knowledge bases~\cite{vrandevcic2014wikidata}
by untrustworthy users;
or deduced from the imprecise answers of random workers on crowdsourcing
platforms~\cite{amsterdamer2013crowd,parameswaran2012crowdscreen}.

Various kinds of uncertainty can hold on the data, which influences our choice of
how to represent it.
The best known is
\emph{fact uncertainty}: we are dealing with statements for which we do not know
whether they are correct or incorrect.
However, there are other situations, such as \emph{order uncertainty}: we are interested in
an order relation on facts (e.g., time, relevance) or on the objects
(e.g., preference, quality), and we only have partial information about this
order (e.g., it was obtained from conflicting user preferences, or by
integrating event sequences that are not synchronized).

The straightforward way to extend existing data management paradigms to uncertain data is to represent
explicitly all possible states of the data (which we call \emph{possible
worlds}), and to define the semantics of queries as returning
all answers that can be obtained on the possible worlds. Of course,
this simple scheme is not practical: there are often exponentially many
possible worlds, so we cannot represent them all, much less query them.
Fortunately, the possible worlds are often \emph{structured}, e.g., by
independence or decomposability assumptions. This encourages us to
design \emph{representation systems}, which concisely describe a
collection of possible worlds, and evaluate queries directly on the
representation, to return a representation of all possible results.

Of course, querying uncertain data implies that, in general, query results will themselves be uncertain.
Still, they have many uses. They allow us to determine whether
some answers are \emph{possible},
or \emph{certain}; or to estimate which ones are \emph{likely}, based on a
probabilistic model on the underlying source of uncertainty (e.g., the
trustworthiness of sources). We can also use them to \emph{specialize} the result of
the query, without reevaluating it from scratch, if we ever obtain information 
that lifts some of the uncertainty. For instance, when we have access to
human users (e.g., via the crowd), we can use the uncertain query results to
estimate which additional knowledge would help reduce the uncertainty, and
ask them the right questions to make the query output more crisp.

We have thus defined semantics for uncertain data. Yet, this does not tell us
whether we can manage it \emph{tractably}. Sadly, in general,
this is not the case. For example, in the
context of fact uncertainty, consider the framework of
\emph{tuple-independent} (or TID) instances~\cite{lakshmanan1997probview}, which
are the simplest kind of probabilistic
relational instances: all facts are independently present or absent with a
given probability. Consider the conjunctive query (CQ) $q : \exists x y \, R(x) S(x, y) T(y)$. It
is \#P-hard~\cite{dalvi2007efficient} to compute the probability that $q$ holds
on an input TID instance, and this is a \emph{data complexity} result,
i.e., it is only in the instance, even when the query is assumed to be fixed.
This contrasts with the $\acz$ data complexity~\cite{abiteboul1995foundations}
of CQs on traditional instances, and makes it necessary in practice to
approximate query results via sampling.
In other contexts, e.g., order uncertainty, or uncertain information
that was partly disambiguated using crowd answers, we do not even
know whether there are good representation systems.

The goal of my PhD is to address this problem from a theoretical angle,
identifying situations where the
\emph{structure} of uncertain data ensures the tractability of exact query evaluation
in terms of possibility,
necessity, and probability. In other words, my goal is to
show that exact query evaluation is tractable when we make assumptions on the
\emph{data}: on the structure of the underlying facts,
on the kind of uncertainty, and on its structure (e.g., fact correlations). The hope
would be to identify tractable classes covering
practical examples of uncertain data, and achieve a theoretical understanding of
why and how we can tractably query them.

The main focus is on fact uncertainty, which is studied in
Section~\ref{sec:presence}. We first study tree representations of
data, in the context of probabilistic XML~\cite{kimelfeld2013probabilistic},
giving examples of how a tractability result for \emph{local} uncertainty models
on trees~\cite{cohen2009running} can be generalized to global uncertainty models
where the \emph{scopes} of uncertain events have bounded overlap. We then move to relational representations, and
explain how the XML tractability results generalize to uncertain relational
instances that have \emph{bounded treewidth}, in the sense of having a
simultaneous bounded-width decomposition of their underlying instance and their
uncertainty annotations. We then describe perspectives to extend this result, the main
one being the problem of reasoning under \emph{uncertain rules}.

My second focus (in Section~\ref{sec:order}) is on order uncertainty. After
motivating this problem, I review our current results of defining a bag
semantics for the positive relational algebra on uncertain ordered data, and
give perspectives such as managing order uncertainty arising from uncertain
numerical values. As a third focus,
Section~\ref{sec:conditioning} studies the question of \emph{conditioning} uncertain
data, e.g., by integrating crowd answers to reduce the uncertainty.
Section~\ref{sec:conclusion} concludes.

\section{Fact Uncertainty}
\label{sec:presence}

\subsection{Trees}
\label{sec:trees}
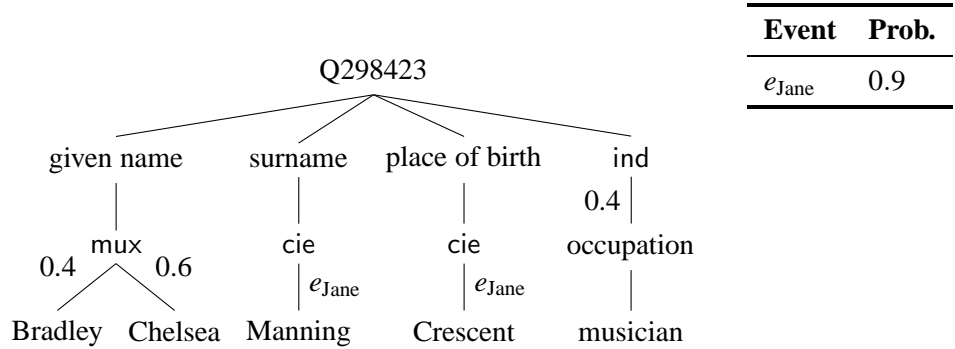
\begin{figure}[t]
\begin{flushright}
  \begin{tabular}{ll}
    \toprule
    {\bf Event} & {\bf Prob.} \\
    \midrule
    $e_{\text{Jane}}$ & $0.9$ \\
    \bottomrule
  \end{tabular}
\end{flushright}
\vspace{-3.2em}
\centering
\begin{tikzpicture}[
  level distance = 3em
]
\Tree
[.Q298423
  [.{given name}
    [.{$\mux$}
      \edge node[auto=right] {0.4};
      [.Bradley ]
      \edge node[auto=left] {0.6};
      [.Chelsea ]
    ]
  ]
  [.{surname}
    [.{$\cie$}
      \edge node[auto=left] {$e_{\text{Jane}}$};
      [.Manning ]
    ]
  ]
  [.{place of birth}
    [.{$\cie$}
      \edge node[auto=left] {$e_{\text{Jane}}$};
      [.Crescent ]
    ]
  ]
  [.{$\ind$}
    \edge node[auto=right] {0.4};
    [.occupation
      [.musician ]
    ]
  ]
]
\end{tikzpicture}
\caption{Example $\prxml$ document}
\label{fig:prxml}
\end{figure}

We start with tree representations of data, i.e.,
XML documents. Figure~\ref{fig:prxml} illustrates such a document (ignoring for
now the annotations): it describes part of the
Wikidata~\cite{vrandevcic2014wikidata} entry about Chelsea Manning.

As the data on Wikidata is not always correct, the information contained in our
tree is \emph{uncertain}; we use the
$\prxml$ probabilistic XML
formalism~\cite{kimelfeld2013probabilistic} to represent this.
For instance, in
Figure~\ref{fig:prxml}, the $\ind$ node describes that the ``occupation''
subtree may or may not be present, with a probability of $0.4$,
\emph{independently} from all other nodes,
modeling our uncertainty about whether this information is correct. The $\mux$
node represents our estimation of the probability that the given name is ``Bradley'' or
``Chelsea''; $\mux$ nodes, unlike $\ind$ nodes, allow choices that are
\emph{mutually exclusive}.

$\ind$ and $\mux$ nodes can represent \emph{local} uncertainty:
indeed, all of their choices have to be taken independently, and only affect
their descendants. As it turns out, query evaluation on trees, in the sense of
determining the probability that a query holds, is tractable under
local uncertainty of this kind~\cite{cohen2009running}, for the usual tree query
languages such as tree-pattern queries or monadic second-order (MSO) queries
without joins.
Tree documents with local uncertainty are thus an example of
\emph{structurally tractable} uncertain data.

However, not all uncertainty sources can be modeled with \emph{local}
uncertainty. Say that the place of birth and surname facts were added to the
Wikidata entry by user Jane.
Rather than modelling them as being independent, we would like to
represent them in a correlated fashion: either user Jane is trustworthy, and both
facts are likely to be true, or she is a vandal, and both are unlikely.
To model this, we use uncertain events, which are a form of \emph{global
uncertainty}. As a simple example, in Figure~\ref{fig:prxml}, the event
$e_{\mathrm{Jane}}$ indicates that we
fully trust Jane with probability $.9$, and the 
$\cie$ nodes (for ``conjunction of independent events'') indicate that the
place of birth and surname facts are either both present or both absent,
depending on whether we trust Jane. As events can be reused at any point in the
document, they can introduce
correlations between arbitrary document parts, so that query
evaluation is generally intractable with events~\cite{kimelfeld2008query}.

This hardness result is not surprising if events are used indiscriminately, but
are there safe ways to use them without leading to intractability?
In~\cite{amarilli2015probabilities} we have answered this
question in the affirmative, by introducing the notion of event
\emph{scopes}, and stating the first (to our knowledge) non-trivial sufficient
condition that guarantees the tractability of query evaluation on
$\prxml$ trees with events. Intuitively,
the scope of an event is the set of nodes where the value of this
event must be ``remembered'' when trying to evaluate a query on the tree; in
Figure~\ref{fig:prxml}, the scope of $e_{\textrm{Jane}}$ are the nodes
``surname`` and ``place of birth'' and their descendants. The scope of a
node~$n$ is the set of events
having $n$ in their scope. We showed that for $\prxml$ documents
where the scope of all nodes have size bounded by a constant, the evaluation of
a fixed MSO query can be performed in PTIME in the input document.

In fact, this claim follows from much more general results about structurally
tractable instances. We now turn to this.

%

\subsection{Tree-Like Data}
\label{sec:treewidth}
\begin{table}
  {
    \setlength{\tabcolsep}{0pt}
    \begin{tabularx}{\linewidth}{X@{\hskip 12pt}X@{\hskip 12pt}r@{\hskip 5pt}c@{\hskip 5pt}l}
    \toprule
    {\bf From} & {\bf To} & \multicolumn{3}{l}{\bf Annotation} \\
    \midrule
    Paris, CDG & Melbourne, MEL & $\mathsf{pods}$ & & \\
    Melbourne, MEL & Paris, CDG & $\mathsf{pods}$ & $\wedge$ & $\neg \mathsf{stoc}$ \\
    Melbourne, MEL & Portland, PDX & $\mathsf{pods}$ & $\wedge$ &
    $\phantom{\neg} \mathsf{stoc}$ \\
    Paris, CDG & Portland, PDX & $\neg \mathsf{pods}$ & $\wedge$ &
    $\phantom{\neg} \mathsf{stoc}$ \\
    Portland, PDX & Paris, CDG & && $\phantom{\neg} \mathsf{stoc}$ \\
    \bottomrule
  \end{tabularx}
  }
  \caption{Example c-instance}
  \label{tab:cinst}
\end{table}

When data cannot easily be represented as a tree, a natural way to write it is
to use relational databases (or instances)~\cite{abiteboul1995foundations}. We can
then represent \emph{uncertain} data using the formalism of
\emph{c-instances}~\cite{ImielinskiL84,green2006models}, which augments
relational instances with propositional annotations on facts using Boolean
events, each event valuation defining a possible world obtained by retaining
only the facts whose annotation evaluates to true.
An example c-instance is given as Table~\ref{tab:cinst}, describing which trips
should be booked depending on the conferences that a researcher wishes to attend: PODS
is taking place in Melbourne and STOC in Portland. We can use 
\emph{pc-instances}~\cite{green2006models,huang2009maybms} to model
probabilistic distributions on instances, simply by giving independent
probabilities to the events of the c-instance.

There are several query languages for relational instances and (p)c-instances:
existentially quantified conjunctions of atoms (known as \emph{conjunctive
queries} or CQs), MSO queries, Datalog~\cite{abiteboul1995foundations}, or some of its
variants such as frontier-guarded Datalog~\cite{baget2011rules}. However, we
know that evaluating a fixed CQ is already \#P-hard in data
complexity, even on TIDs~\cite{lakshmanan1997probview} which are much less
expressive than pc-instances.

Yet, as we saw in the previous section, hardness does not necessarily hold
for tree-shaped data. Could we then show the tractability of query answering on TIDs
which are assumed to be tree-shaped? In fact, we can 
show~\cite{amarilli2015probabilities} that tractability holds for TID instances of
\emph{bounded treewidth}~\cite{robertson1986graph}, which intuitively requires that
they are close to a tree.

\begin{theorem}
\label{thm:tid}
Defining the \emph{treewidth} of a TID as that of its underlying relational
instance (forgetting about the probabilities), for input TIDs with treewidth
bounded by a constant, the evaluation of a fixed MSO query can be performed in
PTIME data complexity. The complexity drops to linear time if we assume
constant-time arithmetic operations.
\end{theorem}

This result cannot directly generalize to pc-tables, because they allow
arbitrary propositional annotations on facts, so CQ evaluation is
\#P-hard in data complexity even on single-fact pc-instances. Hence, to cover
pc-tables as well, we would need to limit the expressiveness of annotations.
Our idea is to write
annotations as Boolean circuits rather than formulae,
and look at the treewidth of the annotation circuit. We can
show~\cite{amarilli2015probabilities} that tractability does \emph{not} follow
from bounded treewidth of the instance and of the circuit in isolation; rather, we must
require the existence of a bounded-width tree decomposition of the instance
\emph{and} circuit,
which respects the link between circuit gates and the facts that they annotate.
We call those bounded-treewidth \emph{pcc-instances}, and we can show:

\begin{theorem}
\label{thm:gen}
Evaluating a fixed MSO query on bounded-treewidth pcc-instances has PTIME
or linear-time data complexity (depending on the cost of arithmetic operations).
\end{theorem}


This general result implies Theorem~\ref{thm:tid} and the scope-based tree tractability
results of the previous section, as these formalisms can be rewritten to
bounded-treewidth pcc-instances.
It relates to Courcelle's theorem~\cite{courcelle1990graph} for usual relational
instances, which shows that
MSO queries (which are generally NP-hard) can be 
evaluated in linear-time data complexity if we assume constant
treewidth. To show this, one compiles~\cite{thatcher1968generalized} the MSO query~$q$, in a data-independent
fashion, to a tree automaton~$A$ which can read tree
encodings of bounded-treewidth instances and determine whether they satisfy $q$. We follow
the same approach, but we show that $A$ can also be run on an
\emph{uncertain} instance~$I$, producing a \emph{lineage circuit} $C$ that describes 
which possible worlds of~$I$ are accepted by $A$. We then show that $C$
has bounded treewidth, and so the probability that $I$ satisfies~$q$ can be
computed from~$C$ via standard message passing
techniques~\cite{lauritzen1988local}. Thus, bounded-treewidth pcc-instances
are \emph{structurally tractable}.

Our method relates to CQ evaluation methods on probabilistic
instances which compute a \emph{lineage} of the query and evaluate the
probability of that lineage. This line of related work has proven fruitful,
e.g., to identify a dichotomy~\cite{dalvi2012dichotomy} between safe and unsafe
\emph{queries} (depending on the data complexity of evaluating them on TID
instances). Our approach is different: we
assume a restriction on the \emph{data}, namely bounded treewidth, and
show that the lineages that we obtain are \emph{always} tractable, for
\emph{any} query that can be compiled to an automaton: beyond CQs, this covers MSO, frontier-guarded Datalog, and more generally guarded
second-order queries. Also, our lineages are circuits rather than formulae, and
are constructed from an automaton for the query rather than an execution
plan. We use this to cast a new light on semiring provenance:
in the case of monotone queries,
our lineage circuits are provenance circuits~\cite{deutch2014circuits} matching
standard definitions of semiring provenance~\cite{green2007provenance} for
absorptive semirings. We show this by connecting the automaton to a new intrinsic
definition of provenance for the query.

Of course, our assumption of bounded-treewidth means that we do not cover many practical use
cases, beyond tree-shaped data. We could address this from a
theoretical angle, as we do not know yet
whether Theorem~\ref{thm:gen} generalizes to weaker assumptions such as
bounded clique-width or hypertree-width~\cite{gottlob2001hypertree}.
However, in more pragmatic terms, we hope to extend our result to \emph{partial} tree
decompositions: we would structure uncertain instances as a
high-treewidth \emph{core} and low-treewidth \emph{tentacles}, and evaluate
queries by combining Theorem~\ref{thm:gen} on the tentacles and
sampling-based approximate methods on the core.
The assumption is that real-world uncertain data, while it may not have bounded
treewidth, should have large low-treewidth parts that can be dealt with using
our exact approach. Approximate query evaluation would then be restricted to the
core, and could thus be made faster or more accurate.
A similar idea (in a more
restricted context) was recently studied
in~\cite{maniu2014probtree}, where it was shown to improve the performance of
source-to-target query evaluation on uncertain graphs.

Another point that we intend to study, in terms of practical applicability, is
the question of combined complexity. Indeed, compiling MSO queries to automata
is generally non-elementary in the query. 
One possibility around this would
be to adapt the construction to monadic Datalog~\cite{gottlob2010monadic};
another one would be to investigate the performance of practical automata compilation
techniques~\cite{henriksen1995mona}.

\subsection{Reasoning Under Probabilistic Rules}
\label{sec:rules}
We conclude our study of structural tractability for tuple uncertainty,
by describing our vision for tractable reasoning under probabilistic rules.

When evaluating queries on incomplete knowledge bases (KBs) such as
Wikidata~\cite{vrandevcic2014wikidata}, we may miss some answers because the
corresponding facts are absent from the KB. However, if we know
some hard constraints about the KB (e.g., the ``located in''
relation is transitive), it makes more sense to say that a query is true if it
is \emph{certain} under the constraints, namely, if it is satisfied by all completions
of the KB that obey the constraints. This is called \emph{open world query
answering}, and it generalizes standard query evaluation (which is the case
where there are no rules).

Our claim is that it would be more useful to reason under soft rules, i.e.,
\emph{probabilistic rules}.
For instance, if the birth date of a person is missing from the KB, we can
deduce a likely range for the date using any other fact about the person. Likewise, a
citizen of a country often lives in that country, and probably speaks the
official language of the country.
Such rules could be produced by association rule mining~\cite{agarwal1994fast},
or using KB-specific methods~\cite{galarraga2013amie}.
Of course, some of the facts that they imply may be wrong,
but on average we expect them to help reduce incompleteness in the KB. Hence, we
would hope to obtain better query answers by asking for the \emph{likely} answers under
many uncertain rules, rather than the certain answers under a few hard rules.

There are already several approaches to reason under uncertain rules, such as
probabilistic programming languages (e.g., ProbLog~\cite{deraedt2007problog}),
or solutions based on Markov Logic Networks~\cite{richardson2006markov} (e.g.,
\cite{jha2012probabilistic}). We would need an approach that satisfies some
desiderata. First, it should be able to express rules which assert the
(probable) existence of \emph{new elements}, or nulls, e.g., a PhD student and
their advisor have probably co-authored \emph{some} paper (which may be unknown to the
KB).
This is not possible with
approaches that focus on vanilla Datalog rules
(e.g.,~\cite{abiteboul2014deduction}), and requires existential Datalog, or
Datalog$^{\smash{+/-}}$, as is done, e.g., in~\cite{gottlob2011conjunctive}.

Second, unlike~\cite{gottlob2011conjunctive}, the approach should be able to
express rules that \emph{usually} apply, not rules which have a certain probability of
\emph{always}
applying. For instance, if we say that citizens of a country are born there with
$80\%$ probability, the semantics of~\cite{gottlob2011conjunctive} is that the rule is either always true or
always false, with probability $80\%$. Our desired semantics is that the rule
applies, on average, in $80\%$ of cases. Maybe closest to our requirements
is~\cite{barany2015declarative}, but the focus of this work is purely declarative,
leaving open the question of the tractability of query answering tasks for such
a model.

Of course, formalizing our desired semantics for probabilistic rules raises many
challenging questions.
First, there may be multiple independent ways to deduce the same fact, so
determining the overall probabilities of new facts is tricky, especially as there
may be correlations, and cyclic derivations where facts are deduced via a path
that involve themselves. Second, the possible consequences of the rules
may be infinite, so that there may be infinitely many possible worlds to
consider (unlike, e.g., pc-instances).
We hope to formalize such a semantics by a variant of the
chase~\cite{abiteboul1995foundations}, yielding both a probabilistic process to
generate possible worlds, and a reasoning process to describe the possible
lineages of facts. Alternatively, another possibility would be eliminate some
rules by rewriting them into the query.

The other challenge posed by probabilistic rules is the question of
tractability. For some languages (e.g., guarded
Datalog~\cite{Gradel:2000:EEM:1765236.1765272} with terminating chase),
we hope to preserve treewidth-based tractability guarantees from the instance to
the rule consequences. If the chase does not terminate, a possibility would be to represent it as a recursive
Markov chain~\cite{benedikt2010probabilistic}, or to truncate it and control the
error.

Beyond guarded rules, it would be practically useful to support 
equality constraints, number restrictions (e.g., ``people have at most
two parents''), or closed-world domains: for instance, when we deduce that a
person has a country of residence, the country probably already exists in the
knowledge base, rather than being a fresh null. However, we do not know which
distribution to assume on such reuses, and we fear that our criteria for
tractability would no longer apply if such reuses are possible.

\section{Order Uncertainty}
\label{sec:order}
We now leave the standard setting of fact uncertainty and move to \emph{order
uncertainty}: we want to model data where we are unsure about the order between
facts or data items. In this setting, to justify the
tractability of uncertain data, we need to invent the right representation
systems to model the uncertain data and the query output.
Of course, uncertain order relations between elements and tuples could in
principle be modeled as fact uncertainty, but this would ignore the \emph{structure} of the
uncertainty: it would create many facts and correlations, leaving little hope
for tractability.

Yet, there are many scenarios where order uncertainty is specifically needed.
For instance, consider the problem of integrating lists of items that are ordered by an
unknown criterion, e.g., a
proprietary relevance function, or the preferences of various
users~\cite{stefanidis2011survey}. If we wish
to take the union of these lists, or to look at pairs (e.g., choices of a hotel
and restaurant in the same neighborhood), there are multiple reasonable choices
to order the result of such operations while respecting the order constraints imposed
by the original lists.  The same problem can arise when integrating logged events
from different machines or files, where the log entries are sequentially ordered
but do not mention a global timestamp (e.g., logs of the fetchmail program, or
\texttt{/var/log/dmesg} on Unix systems); or when integrating concurrent edits
to a document in a version control system~\cite{ba2013uncertain}. The same
problem can occur when searching for the top-$k$ most frequent itemsets in
data mining: if we only have an incomplete view of the data to mine, as 
in our study of data mining on the crowd~\cite{amarilli2014complexity}, we need
to reason under incomplete information about the order relation on the support
value of itemsets.

We have studied this problem and proposed a bag semantics for the positive
relational algebra that applies to relations with uncertain
order~\cite{amarilli2015querying}, which relies on labeled partial orders as its
representation system.
Again, we observe that many tasks on the resulting representations are
intractable to solve: for instance, given a labeled partial order, we cannot
tractably determine whether an input total order is one of its possible worlds.
Yet, for this problem, some specific structures of partial orders are tractable,
such as the ones that were constructed on unordered relations, or totally ordered
relations (depending on the semantics of operators).

Many questions on order uncertainty are still open. For instance, it would be
nice to specify a compositional semantics for the order manipulation operators
of SQL, to formalize all possible reasonable behaviors of SQL implementations. However,
we would need to extend our representation system to more operators, and
to set semantics as well as bag semantics. It would also be interesting to
extend our approach to allow both fact and order uncertainty, for
instance by extending our constructions to support provenance.

Another challenge is to extend our uncertain model to a probabilistic
model, but doing so for order uncertainty is harder than going, e.g., from
c-tables to pc-tables. How can we define a probability distribution on the possible ways to
order the data?
One possibility is to study order that
arises from numerical values (e.g., support, in our data mining scenario).
We have initial ideas~\cite{amarilli2014uncertainty}, but there are a lot of
open questions left. Some are definitional: What are the possible worlds? What is our best guess on how to
interpolate missing numerical values on partially ordered data? Others are operational:
even counting the possible worlds of partially
ordered data may be intractable~\cite{brightwell1991counting}.

\section{Conditioning}
\label{sec:conditioning}
Last, we turn to data that has been
\emph{conditioned}~\cite{tang2012framework}: starting with an original uncertain
data instance, we have revised it to force the outcome of certain probabilistic
events, given new observations or additional information.

The motivation for this kind of uncertain data is very general, because
uncertain data can often be made more certain if we are ready to pay the price.
For instance, we can often ask a human expert
to verify whether a fact is really true, or whether an event 
occurred or not. If we do so, we must figure out two things: which question to
ask, and how to incorporate the answer to our uncertain model.

The answer integration step already poses a problem of tractability: for instance, we
can easily 
condition a c-instance to indicate that an \emph{event} is true, but it
is much harder to force a \emph{fact annotation} to be true, as it can be an
arbitrary formula.
Further, we do not know at all whether structural
tractability guarantees on the original instance can be preserved by
conditioning. We have good hopes for this to be possible, as existing
work in the probabilistic XML context has shown
that it is tractable to query a document that has been conditioned using
a specific language of constraints~\cite{cohen2009incorporating}; note,
however, that this work does not attempt to construct an actual probabilistic
XML document that would represent the distribution obtained by conditioning.

An entirely different issue is to deal with the first step of 
choosing which query to ask.
It is tricky to even define what the best question is, and even harder to find a
sensible definition that is tractable to evaluate.
The most relevant study of this issue may come from crowd data sourcing:
when we try to
extract knowledge from a crowd of human users, we are never sure about what we
know, because we can never fully
trust the answers that have been produced by the crowd workers. Yet, from
our current knowledge and our current estimation of the likely answers, we must
decide what is the next question that we should ask to the
crowd~\cite{amsterdamer2013crowd}, to reduce our uncertainty on the final answer.
To our knowledge, however, existing crowd data sourcing
techniques~\cite{parameswaran2012crowdscreen} use very ad-hoc representations
which are specific to some simple query types.

Hence, it is an important challenge to design a generic uncertainty
representation framework suitable for such iterative
scenarios: at each step, the data is conditioned based on our observations, and
we need to choose the queries that we intend to make, relative to their cost.
Beyond crowdsourcing, we believe that our vision of such a
system~\cite{amarilli2014unsaid} applies to many situations that involve a tradeoff between spending more
resources and acquiring more knowledge.

\section{Conclusion}
\label{sec:conclusion}
We have presented our results about how to deal with order uncertainty, and fact uncertainty on tree and tree-like
instances. We have presented many perspectives to extend
them: for instance, representing the consequences of uncertain deduction rules,
or the result of conditioning the existing data with additional information.

There are interesting directions left to explore. An important one would be to
evaluate the practical
applicability of what we propose,
on datasets or for concrete tasks involving uncertain ordered data
or low-treewidth data.
The design of a practical implementation would also raise theoretical questions:
How to combine our methods with approximate methods such as sampling? Which
optimizations would help us
deal with the high combined complexity?

In terms of representations, we hope to understand how order and fact
uncertainty can be combined, and whether the result could be extended to
cover more uncertainty types, such as the result of conditioning.
Indeed, we believe that a fundamental
challenge for uncertain data representation is to support dynamic situations,
where the data can evolve: new facts are extracted, deduction rules are fired,
and existing information is disambiguated and clarified through human queries or
complex processing. Designing such a framework would be both
a theoretical and a practical challenge.

\paragraph*{Acknowledgements} This work has been supported in part by the
NormAtis project funded by the French ANR. We are grateful to the anonymous reviewers for
their helpful comments.

{
\bibliographystyle{abbrv}
\bibliography{main}
}

\end{document}